\documentclass[doublecol]{epl2}	

\usepackage{graphicx}
\usepackage{amsmath}

\usepackage{color} 

\title{Fluctuations and the role of collision duration in reaction-diffusion systems}

\shorttitle{Fluctuations and the role of collision duration in RD systems}

\author{
  Fernando Peruani\inst{1}\thanks{Email: fernando.peruani@unice.fr} \and
  Chiu Fan Lee\inst{2}\thanks{Email: c.lee@imperial.ac.uk}  }

\shortauthor{Peruani and Lee}

\institute{
\inst{1} Laboratoire J.A. Dieudonn{\'e}, Universit{\'e} de Nice Sophia Antipolis, UMR 7351  CNRS , Parc Valrose, F-06108 Nice Cedex 02, France \\
\inst{2} Department of Bioengineering, Imperial College London, London SW7 2AZ, U.K. \\
 }


\abstract{In a reaction-diffusion system, fluctuations in both diffusion and reaction events, have important effects on the steady-state statistics of the system. 
Here, we argue through extensive lattice simulations, mean-field type arguments, and the  Doi-Peliti formalism  that the collision duration statistics -- i.e., the time two particles stay together in a lattice site --  plays 
a leading role in determining the steady state of the system. We obtain approximate expressions for the average densities of the chemical species and for the critical diffusion coefficient required to sustain the reaction.
}

\begin{document}
\maketitle



Reaction-diffusion 
systems are fundamental in modeling nonequilibrium processes arisen naturally in chemistry, physics, biology and ecology. They also provide a conceptual basis for many phenomena such as pattern formation and wave propagation \cite{cross}. A reaction-diffusion system is typically studied from the perspective of a set of partial differential equations in which fluctuations in particle diffusion and reactions are ignored.  Although such an approach is undoubtedly applicable to a wide range of systems, it has also been realized that when the concentrations of the reacting agents are low, fluctuations can render the deterministic classical picture qualitatively incorrect \cite{redner, lee1995, grima, tauber2005, winkler2012}. To incorporate these fluctuations into the analysis, a number of popular methods have been employed that include
van Kampen's system size expansion \cite{gardiner}, Kramers-Moyal approximation scheme \cite{gardiner}, and the Doi-Peliti (DP) field-theoretic method \cite{tauber2005}. 
{In addition, heuristic approaches have also been successfully applied to study the   $A+B \to C$ reaction-diffusion system~\cite{galfi1988, cheng1989, privman}.}
Among these theoretical tools, approximations as the ones involved in the  van Kampen and Kramers-Moyal method are adequate when the densities of the reacting agents are not too small~\cite{grima}, {while 
the DP method has been employed to study extensively the dynamics of extinction in reaction-diffusion systems \cite{lee1995,oerding2000, tauber2005,reichenbach07a,winkler2012}, and to some degree the corresponding steady state values~\cite{rey1997}}. 
In this paper, we focus on how  fluctuations modify the steady-state picture of the underlying reaction-diffusion system, which is currently under intense investigation \cite{miramontes2002, gonzalez2004, gonzalez2005, peruaniPRL2008,butler2011,woolley2011, belik2011}. Specifically, we contribute to the subject  by investigating numerically and analytically how fluctuations affect the steady state of a chemical reaction with an absorbing phase. We take as example of such type of chemical reaction 
a well-known reaction-diffusion model, the Susceptible-Infected-Recovered-Susceptible (SIRS) epidemic model \cite{murray}.  We begin by introducing the SIRS model and we argue qualitatively why slow diffusion will drive the disease to extinction in the low density limit of the SIRS model.  We then perform extensive lattice simulations in two dimensions to elucidate the relationships between steady-state densities, the diffusion constant, and the infection rate. 
 To gain further physical insight into the behavior of the system, we study analytically how the reaction rate is modified by i) analysing the collision duration of a pair of particles, and ii) by employing an approximation scheme based on the DP method.  We show that both methods lead to qualitatively good agreements with our numerical results. Our findings led us to speculate that the  fluctuations in collision duration are the most relevant mechanism to consider in order to predict the steady-state behavior of a population of reacting chemical species diffusing in space.

\section{SIRS model}\label{sec:SIRS}
We take as example of chemical reaction the SIRS model, which is a standard model for the spreading of an infectious disease. In this model, the total number of agents $N$ is conserved and there are three types of individuals: susceptible ($S$), infected ($I$), and recovered ($R$).  
In terms of chemical reactions, the dynamics of the system can be described by the following scheme:
\begin{eqnarray}\label{eq:reactions}
\begin{array}{ccc}
S+I \stackrel{\alpha}{\rightarrow}  2I , & 
I  \stackrel{\beta}{\rightarrow}       R , &
R \stackrel{\gamma}{\rightarrow} S \, ,
\end{array}
\end{eqnarray}
where $\alpha$, $\beta$, and $\gamma$ are the reaction rates. In the mean-field limit, i.e, when spatial heterogeneity is neglected, the above chemical reaction scheme is soluble
and exhibits two steady states or phases, the so-called absorbing phase, characterized by the extinction of types $I$ and $R$, 
and the active phase, where all three types co-exist. In the active phase, the steady-state density of the susceptible individuals is
\begin{equation}
\label{eq:rho_crit}
\rho_S(t\rightarrow \infty) =\rho^{*}_S = \frac{\beta}{\alpha}
\ ,
\end{equation} 
while the densities for the infected and recovered are $\rho_I^{*}=[\gamma/(\gamma+\beta)][1-\rho^{*}_S]$ and $\rho^*_R=[\beta/(\gamma+\beta)][1-\rho^{*}_S]$ respectively, with 
 $\sum_{k=\{S,I,R\}} \rho^{*}_k=\rho_0$ and $\rho_0$ being the overall density of individuals in the system.
%
The absorbing phase is then defined by Eq.~(\ref{eq:rho_crit}) with $\rho_S/\rho_0=1$. 
Let us now go a bit further and imagine that particles move in a two-dimensional space, with particle motion being Brownian and 
characterized by a diffusion coefficient $D$. 
For pedagogical reasons, let us look at the inadequate (but often useful) conventional reaction-diffusion equations employed to describe these systems \cite{murray}: 
\begin{subequations}
\label{eq:mfa_slow}
\begin{eqnarray}
\label{mfa_s} \partial_t {\rho_S} &=& D \nabla^2 \rho_S -\alpha \rho_I \rho_S
 + \gamma \rho_R \\
\label{mfa_i} \partial_t {\rho_I} &=& D \nabla^2 \rho_I + \alpha \rho_I
\rho_S - \beta \rho_I 
\\
\label{mfa_r} \partial_t {\rho_R} &=& D \nabla^2 \rho_R +\beta \rho_I -\gamma \rho_R
\ .
\end{eqnarray}
\end{subequations}
Note that by performing linear stability analysis of Eq.~(\ref{eq:mfa_slow}) around the absorbing phase (i.e., by setting $\rho_S \approx \rho_0+\epsilon e^{\lambda t + i \mathbf{x} \cdot \mathbf{k}}$, etc), we find the dispersion relation $\lambda = -D k^2 + (\rho_0 \alpha - \beta)$. This means that even though we have added the  diffusion terms, the stability of the absorbing phase does not depend on the diffusion coefficient $D$. This is not what is observed in our simulations, and constitutes one example of the well known phenomenon that  fluctuations can make predictions from conventional reaction-diffusion equations qualitatively incorrect~\cite{redner, lee1995, grima, tauber2005, winkler2012}. 
%

\begin{figure}
\centering
\includegraphics[scale=.25]{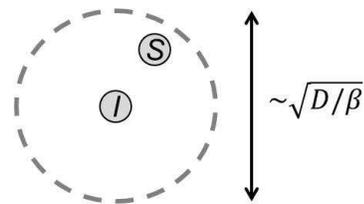}
\caption{(color online) 
An infected agent can roam around an area of $D/\beta$ before recovering. If the density is so low that there is not even one susceptible agent inside this area, then the epidermic will die out irrespective of how high the infection rate is.
}  \label{fig:schematic}
\end{figure}

\begin{figure}
\centering
\resizebox{\columnwidth}{!} {\includegraphics{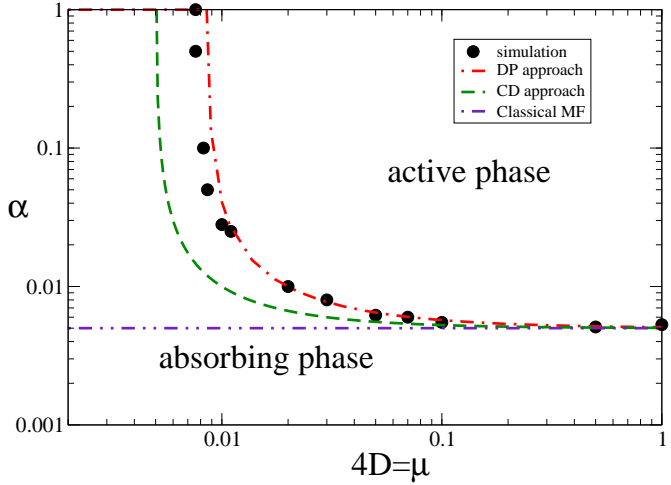}}
\caption{(color online) Phase diagram. The curve separates the active and the absorbing phase. For a given reaction rate $\alpha$, there 
is critical diffusion coefficient $D$ above which the active phase exists. 
Parameters $L=256$, $\rho_0=0.1$, and $\beta=\gamma=5\times 10^{-4}$.}  \label{fig:PhaseDiagram}
\end{figure}

\section{Intuitive argument}\label{sec:intuitive}
We now present an intuitive argument, along the lines of~\cite{redner}, to illustrate why the diffusion constant magnitude should have a drastic effect on determining the phase of the system: 
At the low density limit and close to disease extinction, the infected individuals can be assumed to be far apart as their numbers are small. Each infected individual 
has a life time in the order of $1/\beta$ as they will recover from the disease after this time.  Within such a time frame, the infected agent 
can roam around an area of size $D/\beta$ (see Fig.~\ref{fig:schematic}), and within this area, the number of susceptible agents is $\rho_0 D/\beta$ where we have assumed that $\rho_s \sim \rho_0$ as we are close to disease extinction. For the disease to persist at the steady state, each infected individual has to infect a susceptible individual before recovering. Therefore, even if the infection rate $\alpha$ is infinite, $\rho_0 D/\beta$ has to be of order one. In other words, if $\rho_0 < \beta/D$, one expects that the disease will go extinct irrespective of how high the infection rate $\alpha$ is. This is indeed qualitatively verified by our lattice simulations (Fig.~\ref{fig:PhaseDiagram}) and by our analytical argument (Eq.~(\ref{eq:critical_alpha})). Of course, this simple argument is not enough to  elucidate the full phase diagram of the system and provides, as we show below, an oversimplified picture of the role of $D$. To have a more complete understanding of the phase diagram, we will first rely on lattice simulations.

\section{Lattice model simulations}\label{sec:latticeSim}
In our lattice model, diffusion is modelled as 
 random jumps on a 2D square lattice of linear size $L$ . The jumping rate is denoted by $\mu$, and hence the diffusion coefficient $D$ is $D= \mu \triangle x^2 / (4 \triangle t$). We allow for multiple agents on the same lattice site. The only interaction that occurs among agents in this model is the infection event between an infected and a susceptible agent that occupy the same lattice site. 
Such an event  is characterized by  the transition rate $\alpha$ as indicated in Eq.~(\ref{eq:reactions}). 
 The dynamics of such system of reacting diffusing agents can be rigorously described by a master equation as the one given by Eq.~(\ref{eq:master}), 
\begin{widetext}
\begin{eqnarray}
\label{eq:master} 
\partial_t
P(\hat{\rho}(\mathbf{x}),t)
&=& D \sum_{\phi=\{S,I,R\}} \sum_{\{\mathbf{x}, \mathbf{x}' \}} \left[
  (\rho_{\phi}(\mathbf{x})+1)
  P(\rho_{\phi}(\mathbf{x})+1,\rho_{\phi}(\mathbf{x}')-1) \right.\\
\nonumber
&& \left. -  \rho_{\phi}(\mathbf{x})
  P(\hat{\rho}(\mathbf{x}))
+
(\rho_{\phi}(\mathbf{x'})+1)
  P(\rho_{\phi}(\mathbf{x})-1,\rho_{\phi}(\mathbf{x}')+1) -  \rho_{\phi}(\mathbf{x'})
  P(\hat{\rho}(\mathbf{x}))
\right]\\
\nonumber
&+& \sum_{\mathbf{x}} \left[
\alpha \left[
(\rho_{S}(\mathbf{x})+1)(\rho_{I}(\mathbf{x})-1)P(\rho_{S}(\mathbf{x})+1,\rho_{I}(\mathbf{x})-1) 
-  \rho_{S}(\mathbf{x})  \rho_{I}(\mathbf{x}) P(\hat{\rho}(\mathbf{x}))
\right] \right. \\
\nonumber
&+& \beta \left[ (\rho_{I}(\mathbf{x})+1)
  P(\rho_{I}(\mathbf{x})+1,\rho_{R}(\mathbf{x})-1) 
-  \rho_{I}(\mathbf{x})
  P(\hat{\rho}(\mathbf{x}))\right] \\
\nonumber
&+& \left. \gamma \left[ (\rho_{R}(\mathbf{x})+1)
  P(\rho_{S}(\mathbf{x})-1,\rho_{R}(\mathbf{x})+1) 
-  \rho_{R}(\mathbf{x})
  P(\hat{\rho}(\mathbf{x}))
\right] \right]
\end{eqnarray}
\end{widetext}
where we have defined
$\hat{\rho}(\mathbf{x},t)=\{{\rho}_S(\mathbf{x}),{\rho}_I(\mathbf{x}),{\rho}_R(\mathbf{x})\}$,
the sum $\sum_{\mathbf{x}, \mathbf{x}'}$ runs over all first pairs of nearest
neighbors, and $\sum_{\mathbf{x}}$ over all lattice sites in the system. 
Eq.~(\ref{eq:master}) describes the evolution of the probability of finding
the system in a given configuration $\hat{\rho}(\mathbf{x})$ at time $t$. 
In addition, the equation always deals with an integer number of particles per
node.  Transitions, either involving diffusion or reaction of particles,
are discrete events where only one particle changes its position or internal
state. 

For simplicity, we have set $\triangle x=\triangle t = 1$ in our simulations, which means that $\alpha$, $\beta$, and $\gamma$ represent 
the probability per time step that the corresponding reaction occurs.  
Simulations have been performed according to the following scheme. At every time step, we go first through all particles and with 
probability $\mu$ we move the particle to one of the nearest neighbor lattice site. Secondly, we go through all 
lattice sites where more than one particle is located and evaluate all possible pairs $S$-$I$. With probability $\alpha$ the corresponding reaction is performed.  
Thirdly, we go through all particles $I$ and $R$ and evaluate the transitions $I \to R$ and  $R \to S$ with probabilities $\beta$ and $\gamma$, respectively. 
These three tasks are performed at every time step.  

\begin{figure}
\centering
\resizebox{\columnwidth}{!} {\includegraphics{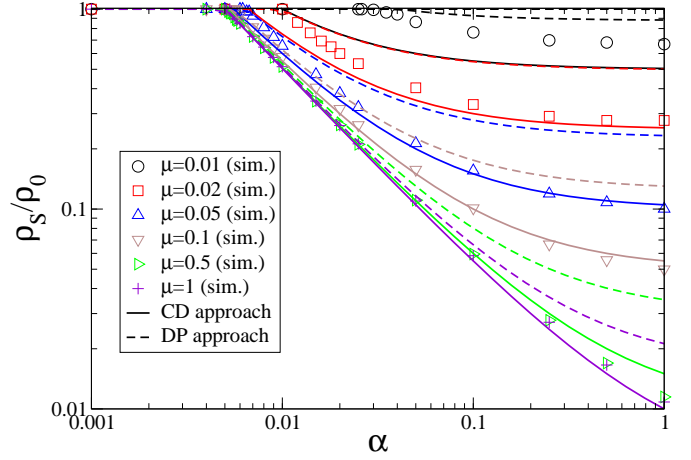}}
\caption{(color online) Steady state value $\rho_S/\rho_0$ as function of reaction rate $\alpha$ for various mobility parameters $\mu$. Parameters $L=256$, $\rho_0=0.1$, and $\beta=\gamma=5\times 10^{-4}$.}  \label{fig:op_vs_alpha}
\end{figure}

\begin{figure}
\centering
\resizebox{\columnwidth}{!} {\includegraphics{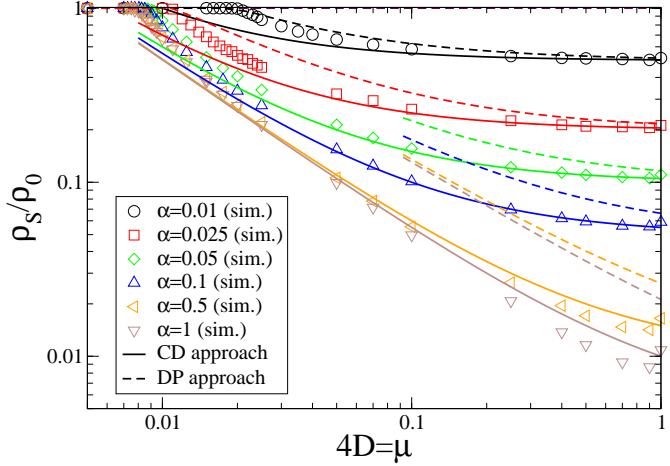}}
\caption{(color online) Steady state value $\rho_S/\rho_0$ as function of the mobility parameter $\mu$ for various reaction rate values $\alpha$. Parameters $L=256$, $\rho_0=0.1$, and $\beta=\gamma=5\times 10^{-4}$.}  \label{fig:op_vs_pmig}
\end{figure}

Figs~\ref{fig:PhaseDiagram}-\ref{fig:op_vs_pmig} show the results of our lattice model simulations. It is evident that the steady-state of the system depends  critically on $D$. We will now attempt to analyze the model analytically with the collision duration (CD) approach.

\section{The collision duration (CD) approach}\label{sec:collisionApproach}
In this section, we expand on our previous intuitive argument in order to  calculate the effective infection rate based on a physical picture of the microscopic dynamics. 
We first consider the typical time required for an infected agent ($I$) to encounter a susceptible agent ($S$), which for diffusional agents can be approximated as $\propto (\rho_S  \mu)^{-1}$. 
We also know that every time that two agents meet  at a lattice site,  the probability that the reaction occurs in a time step is given by 
$\tilde{\alpha}=\alpha \Delta t$. 
Furthermore, the probability that the particle stays at the same lattice site during $\Delta t$ is $\theta = 1-\tilde{\mu}$, with $\tilde{\mu}=\mu \Delta t$. Thus, the probability $p_c(n)$ that it remains 
for $n$ time steps is given by $p_c(n)=\theta^n (1-\theta)$. 
%
On the other hand, the probability $p_r(n)$ that a reaction occurs during the $n$ time steps is $p_r(n)=1- (1-\tilde{\alpha})^{n+1}$. 
In the limit of $\Delta t \to 0$, $p_r(\omega)=1-\exp(-\alpha \omega)$ and $p_c(\omega)=\mu \exp(-\mu \omega)  d\omega$, which means that  $\langle \omega \rangle$ scales with $\mu$ as $\langle \omega \rangle = (\mu)^{-\xi}$, with $\xi=1$. 
%
%
Then, the probability $\psi$ that a reaction occurs during an average  collision event is given by:
\begin{eqnarray}
\label{eq:psi} \psi &=& \sum_{n=0}^{\infty} \theta^n (1-\theta)
\left[ 1- (1-\tilde{\alpha})^{n+1}\right]  \\
\nonumber
&\approx& \int_0^{\infty} d\omega \, \mu  e^{-\mu \omega} \left( 1 - e^{-\alpha \omega}\right) = \frac{\alpha}{\alpha+\mu} \ , 
\end{eqnarray}
where we have used the limit $\Delta t \to 0$ to approximate the sum. 
%
We stress that essentially we are using the residence time distribution to estimate the probability that an encounter results in a successful reaction. 
This time is nothing else than the collision duration.
Moreover, simulations with reactive self-propelled disks (SPD) moving off-lattice can be accurately described by 
using an scheme similar to the one presented here, where $p_c$ is explicitly the collision duration distribution 
which can be empirically measured from the simulations. The collision duration in SPD simulations was found to be exponentially distributed as our estimate for $p_c$~\cite{peruani2012b}. 
Putting together these results, we can express  the rate equation for  $\rho_S$  as
\begin{equation} \label{eq:evol_rhoS}
\partial_t \rho_S = -\psi \mu \rho_S \rho_I +\gamma \rho_R\ .
\end{equation}
This in turn suggests that $\alpha_{eff} = \mu \psi(\alpha,\mu)$.
Inserting this into Eq.~(\ref{eq:rho_crit}), we obtain:
\begin{equation}
\label{eq:asymptotic_rho} \rho^{*}_S = \frac{\beta}{\alpha} \left(1+  \frac{\alpha}{\mu} \right) \  .
\end{equation}
%
Figs.~\ref{fig:op_vs_alpha}  and~\ref{fig:op_vs_pmig} compare Eq.~(\ref{eq:asymptotic_rho}) with simulations data and shows that the CD approach provides a reasonable estimate of the average density of the chemical species with the correct functional dependency with $\alpha$ and $\mu$.
From Eq.~(\ref{eq:asymptotic_rho}), it is possible to obtain the critical reaction rate $\alpha_c$ as function of $\mu$ for a given set of parameters $\beta$, $\gamma$ and $\rho_0$: 
\begin{equation}
\label{eq:critical_alpha} \alpha_c = \frac{\beta}{\rho_0 - [\beta/\mu]} \, .
\end{equation}
%
%
Fig.~\ref{fig:PhaseDiagram} shows that the CD approach even allows us to (roughly) estimate the boundary between the active and absorbing phase. 
{It is worth mentioning that the CD approach provides a correction to the classical mean-field at any spatial dimension.} 

\begin{figure}[t]
\begin{center}
\includegraphics[scale=.43]{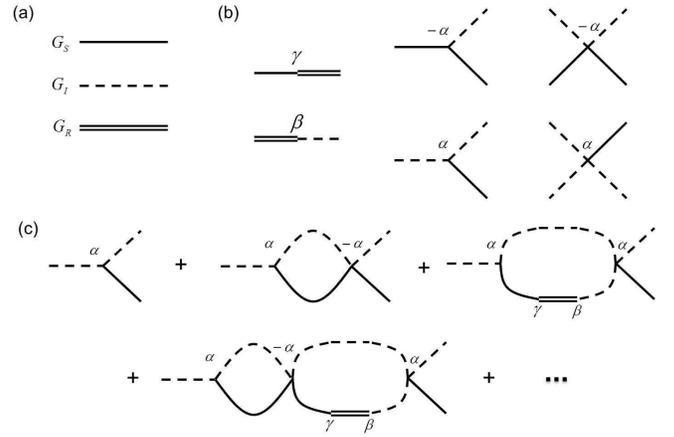}
\caption{(a) The propagators for $S$ and $I$ are denoted by $G_S$ and $G_I$ respectively. (b)  The  diagrams depict the vertices appearing in the action.
(c) The series of sequential-one-loop diagrams captured in the calculations of $\alpha_{eff}$.}
\label{feynman}
\end{center}
\end{figure}

\section{Doi-Peliti Formalism}\label{sec:DoiPeliti}
In the previous section, we have highlighted the importance of the temporal sequence of pairwise interactions between an infected and a susceptible individuals. Here, we will 
employ the well known perturbative DP method to perform calculations on the steady-state densities by focusing again on the temporal sequence of binary interactions. 
{Through the DP formalism we aim to write the evolution of $P(\hat{\rho}(\mathbf{x}),t)$ given by Eq.~(\ref{eq:master}) as a Schr\"{o}dinger-like equation, and then to evaluate average values by making use of a diagrammatic expansion of the resulting path integral formulation. 
The details behind this complex procedure are well documented in the literature, and we refer the interested reader to~\cite{tauber2005}. 
Here, we only outline the main steps to provide, for the non-expert reader, a rough idea about the procedure. 
The first step involves the representation in Fock space of  Eq.~(\ref{eq:master}) that leads to  $|\phi(t)\rangle = \sum_{\hat{\rho}(\mathbf{x})}  P(\hat{\rho}(\mathbf{x}),t) |\hat{\rho}(\mathbf{x})\rangle$. 
The evolution of  $|\phi(t)\rangle$ is given by $\frac{dt|\phi(t)\rangle}{dt}= - \hat{H}|\phi(t)\rangle$, whose formal solution is $|\phi(t)\rangle= \exp(-\hat{H} t) |\phi(0)\rangle$, with $|\phi(0)\rangle$ representing the initial condition. 
The `Hamiltonian' $\hat{H}$ is nothing other than the RHS of Eq.~(\ref{eq:master}) expressed in term of local raising and lowering operators for the different types individuals, which we denote as ${\hat{s}^{\dagger},\hat{i}^{\dagger}, \hat{r}^{\dagger}}$ and ${\hat{s},\hat{i}, \hat{r}}$, respectively. 
Since the diffusive terms are standard~\cite{tauber2005}, we focus on the chemical reaction terms. The reaction $I \to R$ contributes the term  $\left[\hat{i}^{\dagger} \hat{i} - \hat{r}^{\dagger} \hat{i}\right]$ to $\hat{H}$,  $R \to S$ the term $\left[\hat{r}^{\dagger} \hat{r} - \hat{s}^{\dagger} \hat{r} \right]$, and 
 $S+I \to 2I$ the terms $\left[ \hat{s}^{\dagger} \hat{i}^{\dagger} \hat{s} \hat{i} - \hat{i}^{\dagger} \hat{i}^{\dagger}  \hat{s} \hat{i} \right]$. Given $\hat{H}$,  one can employ the Trotter formula together with the coherent states representation to relate the temporal evolution of the  Schr\"{o}dinger equation to  a field theory calculation.  A further customary simplification is the use of the so-called Doi-shift: $s^{\dagger} = 1 + s^{*}$,  $i^{\dagger} = 1 + i^{*}$, and $r^{\dagger} = 1 + r^{*}$ in the action  $\hat{S}$, whose final form is shown in Eq.~(\ref{eq:action}).} 
\begin{widetext}
\begin{eqnarray}
\label{eq:action} {}
\hat{S}(\{\mathbf{\phi}, \mathbf{\phi}^*\};0,t)=\int^t_0 dt' \int d^dx \left[ s^* (\partial_t - D \nabla^2) s +
  i^* (\partial_t - D \nabla^2 +\beta ) i + r^* (\partial_t - D \nabla^2
  +\gamma ) r \right. \\
\nonumber
\left. -\gamma s^* r - \beta r^* i + \alpha (s^* s i - i^* s i + s^* i^* s i - i^*
i^* s i ) - ({\rho}_{S_0} s^* + {\rho}_{I_0} i^* + {\rho}_{R_0} r^*)
\delta(t) \right] \ .
\end{eqnarray}
\end{widetext}

In Eq.~(\ref{eq:action}), 
${\rho}_{S_0}(\mathbf{x})$, ${\rho}_{I_0}(\mathbf{x})$,
${\rho}_{R_0}(\mathbf{x})$ correspond to the initial numbers of agents per lattice site for the corresponding types. 
To calculate the evolution of the density of the $\omega$-type where $\omega =\{ s,i,r\}$, we can employ the following formula:
\begin{eqnarray}
\label{eq:back_to_density} 
\rho_\omega(\mathbf{x},t) = \frac{ \int D \{ \mathbf{\phi}, \mathbf{\phi}^*\} \,\omega\,
  e^{-\hat{S}(\{\mathbf{\phi},\mathbf{\phi}^*\};0,t)} }{ \int D \{ \mathbf{\phi}, \mathbf{\phi}^*\}  e^{-\hat{S}(\{\mathbf{\phi},\mathbf{\phi}^*\};0,t)}} \,,
\end{eqnarray}
where  $\int D \{ \mathbf{\phi}, \mathbf{\phi}^*\} \hdots$ represents functional integrations over the fields $\{s,i,r,s^*,i^*,r^*\}$ { and $\omega$ has to be replaced by $s$, $i$ or $r$}. Note that if we ignore the quartic terms in the action in Eq.~(\ref{eq:action}), the above field theory reduces---by finding the extrema of $\hat{S}$, i.e., $\frac{\delta \hat{S}}{\delta s^*}=\frac{\delta \hat{S}}{\delta i^*}=\frac{\delta \hat{S}}{\delta r^*}=0$---to the ``classical'' reaction-diffusion equations shown in Eq.~(\ref{eq:mfa_slow}).
In physical terms, the quartic terms in the action accounts for the fluctuations inherent in the diffusion and infection events that are ignored by the  ``classical'' equations  \cite{tauber2005}.

In principle, the steady state configuration of the system can now be obtained by evaluating the functional integrals in Eq.~(\ref{eq:back_to_density}). In practice, such a calculation is unfortunately intractable and thus approximation schemes are needed. Motivated by the results of the previous section, we will employ  
the Feynman diagram method to keep track of the temporal sequence of pairwise interactions between an infected and a susceptible individuals. 
In this graphical method (Fig.~\ref{feynman}), the propagators for the three types of agents are denoted by $G_\omega({\bf k},t)$. For instance,
\begin{eqnarray} \label{eq:propagator}
G_I({\bf k},t) &=& \Theta(t) e^{-(\epsilon(k)+\beta)t} \ ,
\end{eqnarray}
where {$ \Theta(t)$ is the Heaviside function, with $ \Theta(t)=1$ for $t\geq0$ and 0 otherwise, and}  
\begin{equation}
\label{eq:epsilon}
\epsilon(k)=4D\sum_{j=1}^2\sin^2(k_j/2)
\end{equation}
 represents the dispersion relation for the diffusion operator on a 2D square lattice~\cite{machado2010,winkler2012}. Interactions between the agents are then captured by the vertices depicted in Fig.~\ref{feynman}(b) that connect the propagators. An exact calculations of Eq.~(\ref{eq:back_to_density}) amounts to summing all of the relevant diagrams. {As mentioned before, motivated by our collision duration 
 analysis, 
 we will only  focus on the sequential pairwise interactions by selectively summing the series of all sequential-one-loop diagrams  depicted in Fig.~\ref{feynman}(c).} Due to the choice of the diagrams considered, the effective infection rate can be expressed as
 a  power series in $\alpha$: {
 \begin{equation}
 \label{eq:aeff2}
 \alpha_{eff} = \alpha+(w_2-w_1)\alpha^2 +(w_1^2 -2w_1w_2+w_2^2)\alpha^3 +{\cal O}(\alpha^4) 
 \ ,
 \end{equation}
where $w_1$ denotes the contribution from the second  diagram in Fig. \ref{feynman}(c) and $w_2$ denotes the contribution from the third  diagram in Fig. \ref{feynman}(c). For instance, 
\begin{eqnarray} \nonumber
w_1 &=& \frac{2}{(2\pi)^2}\int_0^\infty d t \int_\Lambda d^2kG_S({\bf k} ,t) G_I(-{\bf k} ,t)
\\
 \label{eq:w1}
&=& \frac{2}{(2\pi)^2} \int_\Lambda d^2k \frac{1}{2 \epsilon(k) +\beta}
\ ,
\end{eqnarray}
where $\Lambda$ denotes the first Brillouin zone: $]-\pi, \pi ]^2$ \cite{machado2010}.
Note that due to the presence of the decay rate $\beta$, the above loop integral is both infrared and ultraviolet convergent and thus $w_1$ can be easily computed numerically. The same applies to $w_2$. Since the sum in Eq.~\ref{eq:aeff2} is a sum of a geometric series, $\alpha_{eff}$ can be rewritten as}
\begin{equation}
\label{eq:eff_a}
\alpha_{eff} = \frac{\alpha}{1+\alpha(w_1-w_2)} \ .
\end{equation}
Now, due to the presence of the diffusion constant in $\epsilon(k)$ (see Eq.~(\ref{eq:epsilon})), $\alpha_{eff}$ also becomes $D$ dependent. In other words, distinct from the mean-field prediction, the critical infection rate that separates the absorbing and active regions depends on the diffusion coefficient $D$. Substituting $\alpha_{eff}$ from Eq.~(\ref{eq:eff_a}) into Eq.~(\ref{eq:rho_crit}), we obtain 
that:
\begin{equation}
\label{eq:asympt_r_DP}
\rho^{*}_S = \frac{\beta}{\alpha} \left( 1 + \alpha (w_1-w_2) \right)\, .
\end{equation}
 Furthermore, 
the phase boundary separating the absorbing and active phases can be obtained by estimating $\rho_S/\rho_0 = 1$, from which we find 
that the critical reaction rate:
\begin{equation}
\label{eq:critical_alpha_DP}
\alpha_c = \frac{\beta}{\rho_0 - \beta (w_1-w_2)} \, .
\end{equation}
%
%
Comparing Eqs. (\ref{eq:asymptotic_rho}) and  (\ref{eq:asympt_r_DP}), and  (\ref{eq:critical_alpha}) and  (\ref{eq:critical_alpha_DP}), it becomes 
evident that both methods, i.e., the CD and DP approach, lead to the same functional prediction, with $(w_1-w_2)$ being approximated by $\mu^{-1}$ in the CD approach. 
In the limit of infinitely fast diffusing particles, both approaches reduce, as expected, to the mean-field prediction, Eq. (\ref{eq:rho_crit}).  
{Notice that the DP equations are derived for multiple occupancy at each lattice site. Given the low density regime we are considering here, we do not think that the corresponding single occupancy version would affect our results. This is also supported by the fact that  a similar macroscopic behavior has been observed in recent simulations with volume exclusion~\cite{peruani2012b}.}

Fig.~\ref{fig:PhaseDiagram} shows good quantitative agreement between the phase 
diagram $\alpha$-$D$ obtained in simulations and by implementing the CD and DP formalisms. 
Figs. \ref{fig:op_vs_alpha}--\ref{fig:op_vs_pmig} show that the DP method also qualitatively agrees with the numerical simulations, and even provides a quantitative description for small values of the $\alpha$ and $D$.  
The (approximated) DP method seems to deviate from numerical results as $\alpha$ and $D$ become large (see Figs. \ref{fig:op_vs_alpha}--\ref{fig:op_vs_pmig}), while the CD approximation provides a more accurate description. 
In short, the CD method provides a better description of the system dynamics far away from the critical point (Figs. \ref{fig:op_vs_alpha} and \ref{fig:op_vs_pmig}), while the (approximated) DP approach predicts more accurately the behavior and position of the critical point (Fig. \ref{fig:PhaseDiagram}).

\section{Concluding remarks}\label{sec:concluding}
A chemical reaction among moving reacting particles (on a lattice) can be exactly described by a master equation as Eq.~(\ref{eq:master}). 
Unfortunately, the direct integration of such equations is typically infeasible. 
Given the complexity of the problem, neglecting fluctuations  seems to be the intuitive way to tackle the problem. 
However, we have seen that a classical reaction diffusion approach, as shown in Eq.~(\ref{eq:mfa_slow}), 
provides a qualitative incorrect description of the problem close to, and far away from, the critical point.  
In order to obtain a physically accurate description of the dynamics, fluctuations have to be, somehow, incorporated in the theory. 
One option is to use the machinery of field theory through the DP formalism.  
{Such a formalism is able to account for all kind of fluctuations occurring in the system that include 
 those resulting from the re-encounters of the infected and susceptible agents as well as their spatial anti-correlation.}      
There is, however, a practical problem that ultimately represents a drawback of the DP approach. The DP formalism provides an exact and reliable description of the problem as long as 
we are able to sum all the involved Feynman's diagrams, but this is in most cases an unfeasible task.  
Moreover, we cannot systematically improve our description since we cannot know which  diagrams contain the leading contribution. 
As a result, we are forced to use our intuition to select a collection of diagrams with the hope of picking up the most relevant ones. 
Figs~\ref{fig:PhaseDiagram}-\ref{fig:op_vs_pmig} show that the diagrams we have selected are enough to provide a correct qualitative description of the 
population dynamics, with the critical point and the steady state population values exhibiting the correct functional dependency with particle diffusion. 
Nevertheless, the quantitative mismatch is evident, particularly far away from the critical point.  This can be due to the uncontrolled approximation scheme we employ, i.e., the set of diagrams we chose to consider. 
On the other hand, with the collision duration approach, we focused on the effect of local,  mobility-induced fluctuations on the population dynamics and   
 showed that 
 this simple approach provides an excellent semi-quantitative description of the long term system behavior, with 
the critical point and asymptotic population values exhibiting the correct functional dependency with particle diffusion. 
{The CD approach provides corrections to the classical mean-field at any dimension.}  
Our findings strongly suggest  that collision duration fluctuations play a leading role in reaction-diffusion models of population dynamics.


\acknowledgments
We acknowledge support by the Max Planck Institute for the Physics of Complex Systems (Dresden), and the Lab. J.A. Dieudonn{\'e} (Nice).


\end{document}